\documentclass[12pt]{iopart}
\usepackage{iopams}
\usepackage{graphicx}

\newcommand{\eqref}[1]{(\ref{#1})}
\newcommand{\commento}[1]{}

\newcommand{\vR}{{R}}

\newcommand{\bx}{{\bf x}}
\newcommand{\bX}{{\bf X}}
\newcommand{\bv}{{\bf v}}

\newcommand{\bu}{{\bf u}}

\graphicspath{{Figures/}}

\begin{document}

\title[How does one compare correlation functions and responses?]{The
fluctuation-dissipation relation: how does one compare correlation
functions and responses?}

\author{D. Villamaina$^{1}$, A. Baldassarri$^{1,2}$, A. Puglisi$^2$ and A. Vulpiani$^{3,2}$}
\address{$^1$ Dipartimento di Fisica, Universit\`a La Sapienza, p.le Aldo Moro 2, 00185 Roma, Italy}
\address{$^2$ CNR - INFM Statistical Mechanics and Complexity Center, p.le Aldo Moro 7, 00185 Roma, Italy}
\address{$^3$ Dipartimento di Fisica and INFN, Universit\`a La Sapienza, p.le Aldo Moro 2, 00185 Roma, Italy}

\ead{dario.villamaina@roma1.infn.it, andrea.baldassarri@roma1.infn.it, andrea.puglisi@roma1.infn.it and
angelo.vulpiani@roma1.infn.it}

\begin{abstract}
We discuss the well known Einstein and the Kubo Fluctuation
Dissipation Relations (FDRs) in the wider framework of a generalized
FDR for systems with a stationary probability distribution.  A
multi-variate linear Langevin model, which includes dynamics with
memory, is used as a treatable example to show how the usual relations
are recovered only in particular cases. This study brings to the fore
the ambiguities of a check of the FDR done without knowing the
significant degrees of freedom and their coupling. An analogous
scenario emerges in the dynamics of diluted shaken granular
media. There, the correlation between position and velocity of
particles, due to spatial inhomogeneities, induces violation of usual
FDRs.  The search for the appropriate correlation function which could
restore the FDR, can be more insightful than a definition of
``non-equilibrium'' or ``effective temperatures''.
\end{abstract}

\maketitle

\section{Introduction}










The idea of a link between dissipation and fluctuations dates back to
Einstein with his work on the Brownian motion~\cite{E56} and his
relation between mobility (which is a non-equilibrium quantity) and
diffusion coefficient (which is an equilibrium quantity). Later
Onsager~\cite{O31,O31b} with the regression hypothesis and
Kubo~\cite{K57,K66}, with linear response theory, investigated in a
deep way the issue of the fluctuation-dissipation relation (FDR). In
the last decades there has been a renewed interest in this topic, see
the contributions of Morris, Evans, Cohen, Gallavotti and
Jarzynski~\cite{ECM,GC,CJ} to cite just some of the most well known
attempts (for a recent review, see~\cite{BPRV08}).

The FDR theory was originally introduced in the Hamiltonian systems
near thermodynamic equilibrium. However it is now clear that a
generalized FDR exists, under very general assumptions, for a large
class of systems with a ``good statistical behaviour'', i.e. with a
relaxation to an invariant (smooth) probability distribution. We
stress that such a condition is quite common, e.g. in any system with a
finite number of degrees of freedom whose evolution rules include some
randomness (for instance non-linear Langevin equations). Unfortunately
the explicit form of generalized FDR depends on the shape of the
invariant probability distribution (which is typically unknown), however
this is only a technical difficulty without conceptual
consequences~\cite{BPRV08}.

We are not concerned with systems without a stationary probability
distribution, e.g. systems showing aging~\cite{BCKM98,MPRR98} and
glassy behaviour~\cite{LN07}. On the contrary, we consider here
 systems with an invariant probability distribution 
satisfying the hypothesis for the validity of the generalized FDR mentioned
before.  For such systems, in our opinion, there is some confusion
about what form of FDR has to be expected. 
 When couplings between the chosen observable and other degrees of
freedom are ignored, the wrong FDR is expected, i.e. the response
function is compared to the wrong correlation: this leads to what is
often called a ``violation'' of FDR.

The structure of the paper is the following: at first, in Section 2,
we discuss the different kinds of FDR obtained in the statistical
mechanics framework.
In Section 3 we
analyze a one-dimensional Langevin equation with memory, which can be
mapped to a multivariate Langevin equation without memory: this
example can be worked out analytically and well illustrates all the
main points of our discussion.  In section 4 some numerical results on
a driven granular gas model are discussed to support our general
discussion with physical examples. Section 5 is devoted to concluding remarks.

\section{Linear response in statistical mechanics}

Let us briefly recall three different kinds of Fluctuation Dissipation
Relations (FDR), commonly used in statistical mechanics when a small
impulsive perturbation is applied to a stationary system. These three
formulae share the feature of relating the system's linear response to
an {\em appropriate} two-time correlation computed in the unperturbed
system. Anyway these relations have different fields of application
and must be adapted with care. For a more pedagogical introduction, we
first set the notation, and then we discuss the three FDR versions.

\subsection{Linear response functions}

We adopt the following notation: $\delta(t)$
denotes the Dirac delta function, and $\delta_{ij}$ the Kronecker
delta,  we use the overline $\overline{\;\cdot\;}$ for non-stationary
averages over many realizations and $\langle \cdot \rangle$ for
averages using the unperturbed stationary probability in phase-space
(assuming ergodicity, this is equivalent to a time-average over a long
trajectory). Accordingly, we use the shorthand notation
\begin{equation}
C_{AB}(t)=\langle A(t)B(0)\rangle
\end{equation} 
to denote the two-time correlation function between observables
$A(\bX(t))$ and $B(\bX(t))$, with $\bX(t)$ the state of the system at
time $t$.  Let us introduce the matrix of linear response functions, whose $ij$ element
reads
\begin{equation} \label{res_def}
\vR_{X_iX_j}(t)\equiv\frac{\overline{\delta X_i}(t)}{\delta
X_j(0)},
\end{equation}
i.e. the mean response of the variable $X_i$ at time $t$ to an
impulsive perturbation applied to a variable $X_j$ at time $0$. If the dynamics of the system is
given, for instance in the form $\frac{d\bX(t)}{dt}={\bf f}(\bX)$, the
mean linear response of the $i$-th degree of freedom to a small
perturbation of the $j$-th component of the vector field $f_j \to f_j
+ \delta f_j$ can be expressed as
\begin{equation} \label{risp_f}
\overline{\delta X_i}(t)=\int_{-\infty}^t ds R_{X_{i}X_{j}}(t-s) \delta f_j(s).
\end{equation}
The case of an impulsive perturbation at time $0$, $\delta f_j = \delta
X_j(0) \delta(t)$, gives back the definition~\eqref{res_def}.\\
Let us also consider the historically important case of a Hamiltonian
system at thermal equilibrium: in this case the perturbation is
typically defined on the Hamiltonian, i.e. $\mathcal{H} \to
\mathcal{H}+\delta \mathcal{H}$, with $\delta \mathcal{H} \equiv
-\delta h(t)B(\bX)$. A linear response function of an observable $A$  to an impulsive field $\delta h (0) \delta(t)$ at time $t=0$, $R_{Ah}(t)=\frac{\overline{\delta A(t)}}{\delta h (0)}$ determines 
the behaviour of $\overline{\delta A}(t)$ for a generic small perturbation $\delta h (t)$:
\begin{equation} \label{risp_h}
\overline{\delta A}(t)=\int_{-\infty}^t ds R_{Ah}(t-s) \delta h(s).
\end{equation}
There are not conceptual differences between the two procedures to
perturbe the state (i.e. with a $\delta X_{i}(0)$ or the introduction
of an extra term in the Hamiltonian): we can simply consider $A$ and
$B$ as two variables of the system. For instance, consider the case
where $\bX$ is the position of a colloidal particle evolving according
to
\begin{equation} \label{transl}
\dot{X}_i=-\frac{1}{\gamma} \frac{\partial H}{\partial X_i}+\sqrt{\frac{2T}{\gamma}} \eta_i,
\end{equation}
with $\eta_i$ independent normalized white noises, i.e. Gaussian
processes with $\langle \eta_i \rangle=0$ and $\langle
\eta_i(t)\eta_j(t')\rangle=\delta_{ij}\delta(t-t')$. Here, the effect
of a perturbation $\Delta \mathcal H = -X_j \delta h(t)$ is equivalent
to a $\delta f_{i}=\delta_{ij}\frac{\delta h(t)}{\gamma}$ and
therefore, from a comparison of Eqs.~\eqref{risp_f} and~\eqref{risp_h},
one has $R_{X_ih}=R_{X_i X_j}/\gamma$.

\subsection{Three different Fluctuation Dissipation Relations}

We can now discuss the
three forms of FDR we are interested in:

\begin{enumerate}

\item the generalized FDR, denoted GFR in the following
\begin{equation} \label{ger}
\vR_{X_iX_j}(t)=C_{X_iS_j}(t),
\end{equation}
where $S_j$, see below, depends on the invariant distribution density in phase space.
This relation is valid (under quite general assumptions,
see~\cite{a72,dh75,ht77,ht82,FIV90,BPRV08}) in a  dynamical
system, whose state is completely determined by the phase
space coordinate $\bX$, and whose dynamics induces an invariant
measure of phase-space $\rho(\bX)$~\footnote{more precisely one assumes
that $\rho(\bX)$ is a smooth non-vanishing function. This condition
surely holds if some noise is included in the dynamics.}, and
\begin{equation}
S_j=-\frac{\partial \log \rho(\bX)}{\partial X_j}
\end{equation}
so that, the correlation reads:
\begin{equation}
C_{X_iS_j}(t)=-\left\langle X_i(t)\frac{\partial \log \rho(\bX)}{\partial X_j}\right\rangle;
\end{equation}

\item the so-called Einstein relation, in the following referred as
  EFR~\footnote{for simplicity we adopt the name ``Einstein relation''
    which, in the literature, has been typically used to denote the
    time-integral of relation~\eqref{er}, e.g. the formula $\mu=\beta
    D$ relating the mobility $\mu$ to the self-diffusion coefficient
    $D$. },
 \begin{equation} \label{er}
R_{AA}(t)\equiv\frac{\overline{\delta A}(t)}{\delta A(0)}=\frac{C_{AA}(t)}{C_{AA}(0)}.
\end{equation}
The most known example of the above relation is given by a Brownian colloidal
particle diffusing in an equilibrium fluid at temperature $T$.
Such a system is described by a linear Langevin equation
\begin{equation} \label{la_free}
m \dot{v} = -\gamma v + \sqrt{2\gamma T}\eta
\end{equation}
where $\eta$ is a normalized white noise, i.e. a Gaussian process with
$\langle \eta \rangle=0$ and $\langle
\eta(t)\eta(t')\rangle=\delta(t-t')$.  It is easy to show that after a
small impulsive force $\delta h(0)\delta(t)$ at time $0$, the velocity
perturbation, which at $t=0$ is $\delta v(0) = \delta h(0)/m$, decays
as $\overline{\delta v(t)}=\delta v(0) \e^{-\gamma t}$.  On the other
hand, since $C_{vv}(t)=C_{vv}(0)\e^{-\gamma t}$, where
$C_{vv}(0)=\langle v^2 \rangle=T/m$, the EFR~(\ref{er}) holds, with $A
\equiv v$ (see~\cite{KTH91}).  The Green-Kubo relations, relating
transport coefficients to the time-integral of unperturbed
current-current correlations are the extension of relation~\eqref{er}
to generic transport processes;

\item the classical Kubo relation, hereafter denoted as KFR,
\begin{equation} \label{kubor}
R_{Ah}=-\beta \frac{d}{dt} C_{AB}(t).
\end{equation}
This relation holds - for instance - when a Hamiltonian system, whose
statistics is described by the canonical ensemble with temperature
$T=1/\beta$, is perturbed by a Hamiltonian
variation $\delta \mathcal{H}=-\delta h(t) B$, which defines
the perturbing force $h$ and the corresponding {\em conjugate field}
$B$ (see~\cite{K66,BPRV08}). Relation~\eqref{kubor} also holds for
Langevin equations with a gradient structure, e.g. Eq.~\eqref{transl}. 
Introducing the quantity $\chi_{AB}(t)=\int_0^t R_{Ah}(t') dt'$~\footnote{For this last quantity we use a
notation where the second subscript $B$ directly refers to the
observable conjugate to the perturbed field $h$: even if it is not
self-evident, this has the advantage of being coherent with the
notation widely used in the literature.},
from~\eqref{kubor} one has
\begin{equation} \label{kuborint}
\chi_{AB}(t)=\beta C_{AB}(0)- \beta C_{AB}(t).
\end{equation}
In some literature~\cite{CK00,ZBCK05,CR03} a deviation
from~\eqref{kuborint} is indicated as a failure of FDR, a mark
of being far from equilibrium, and is used to define new
``non-equilibrium'' temperatures.

\end{enumerate}

As the name suggests, the GFR (iii) includes both EFR (i)
and KFR (ii) for some choices of the dynamics or of the
stationary distribution $\rho(\bX)$~\cite{BPRV08}.  This can be shown for the case
$A=X_i$ and $B=X_j$ and is easily generalized to any other case, as discussed above:

\begin{itemize}

\item GFR $\to$ EFR: this happens when the invariant distribution in
  phase space is of the form  $\rho(\bX)=\exp(-\frac{1}{2T}\sum_i X_i^2)/Z$ (being $Z$ a normalizing
  constant), then one has $S_j=X_j/T$ and it is immediate to obtain
  EFR starting from relation GFR;

\item GFR $\to$ KFR: this happens for Hamiltonian systems in the
canonical ensemble, or Langevin equations with gradient structure when
a small force is applied: in both cases one has $\rho(\bX)=\exp[-\beta
\mathcal{H}(\bX)]/Z$, so that the GFR involves the quantity
$S_j=\beta\frac{\partial \mathcal{H}}{\partial X_j}$. If the dynamics
is given, for instance, by an overdamped Langevin equation of the
kind~\eqref{transl}, one has that $S_j=-\beta\gamma \dot{X}_j$,
i.e. $R_{X_i X_j}=\beta \gamma \langle X_i(t) \dot{X}_j(0)\rangle$
and, considering the discussion after Eq.~\eqref{transl}, the KFR
$R_{X_i h}=\beta \langle X_i(t) \dot{X}_j(0)\rangle$ is immediately
derived.

\end{itemize}

\section{Langevin equation with memory} \label{linlan}

Let us now consider a system that does not necessarily fall neither
under the hypothesis of EFR, neither under those of KFR. Our choice
here goes to a linear stochastic equation with memory:
\begin{equation}
m\ddot{x}=-kx-\int^{t}_{-\infty}\Gamma(t-t')\dot{x}(t')dt'+\eta(t) \label{3mainequation}
\end{equation}
where $m$ is the mass of the tracer, $k$ is the constant of an elastic
force, $\Gamma(t-t')$ is the friction kernel, $\eta(t)$ is a stochastic
force acting as a thermostat. 

When there is no memory, i.e. $\Gamma(t-t')=2 \gamma \delta(t-t')$ and
$\left<\eta(t)\eta(t')\right>=2 \gamma T\delta(t-t')$, the
usual Langevin equation is recovered, leading to EFR in the case $k=0$, Eq.~\eqref{la_free}, or KFR in the overdamped case, Eq.~\eqref{transl}.

For a generic memory kernel, Kubo~\cite{K66} has also shown that EFR or KFR are recovered, provided that
\begin{equation} \label{eq_cond}
\left<\eta(t)\eta(t')\right>=T\Gamma(t-t'). \label{3equilibriumfdt}
\end{equation}

Here we consider a memory kernel of the kind:
\begin{equation}
\Gamma(t-t) = 2 \gamma_f \delta(t-t') + \frac{\gamma_s}{\tau_s}\exp\left(-\frac{t-t'}{\tau_s}\right)
\end{equation}
and a noise $\eta(t)=\rho_{f}(t)+\rho_{s}(t)$ where $\rho_f(t)$ and
$\rho_s(t)$ are two independent Gaussian processes with zero means and
\begin{equation}
\left<\rho_{f}(t)\rho_{f}(t') \right>=2 T_f \gamma_f \delta(t-t') \phantom{mm} \left<\rho_{s}(t)\rho_{s}(t') \right>=T_{s} \frac{\gamma_{s}}{\tau_{s}}e^{-\frac{\vert t-t' \vert}{\tau_{s}}}. 
\end{equation}
so that condition~\eqref{eq_cond} is recovered when $T_f=T_s\equiv T$
(``s'' and ``f'' subscripts stay for ``slow'' and ``fast''
respectively). In general, however, $T_f \neq T_s$. Note also
that $T_f$ and $T_s$ have the dimension of a temperature: indeed the
model in its overdamped limit has been proposed as
an example of system coupled to two different baths acting on
different time-scales~\cite{CK00,ZBCK05}.



\subsection{A Markovian equivalent model}

The first observation about the system in study is that, because of
the memory term, its dynamics after time $t$ cannot be deduced by the knowledge of
$x$ and $v$ at time $t$: in fact, the evolution depends on its
history, i.e. the dynamics is non-Markovian, and the GFR cannot be directly applied.

However, it is possible to recover Markovianity, at the price of
adding additional degrees of freedom.  In other words, it is possible (and it will be done in the next section)
to recast equation~(\ref{3mainequation}) in a linear,
multi-dimensional, Langevin equation:
\begin{equation}
\frac{d\bx}{dt}=-A{\bf x}+ {\bf \phi}, \label{2main}
\end{equation}
where $\bx$ e $\bf{\phi}$ are N-dimensional vectors and $A$ is a real
$N\times N$ matrix, in general not symmetric.
In addition, now $\bf {\phi}(t)$ is a Gaussian process,
with covariance matrix:
\begin{equation}
\left<\phi_{i}(t')\phi_{j}(t)\right>=\delta(t-t')D_{ij},
\end{equation}
and the real parts of $A$'s eigenvalues are positive.
The stationary probability density is~\cite{R89}:
\begin{equation}
P(\bx)=(2\pi)^{-N/2}[\mathrm{Det} \sigma]^{-1/2}\exp\left\{-\frac{1}{2}\sum_{ij}\sigma^{-1}_{ij}x_i x_j \right\},
\end{equation}
where $\sigma$ is a symmetric matrix determined by the following  relation:
\begin{equation}
D= A\sigma+\sigma A^{T}. \label{2sigmadefinition}
\end{equation}

We can now explicitly study the fluctuation and response properties of
the system since the dynamics, being now Markovian, satisfies the
hypothesis of applicability of GFR. First, we recall the definition of
correlation matrix $C_{ij}(t)=\langle x_{i}(t)x_{j}(0)\rangle$,
in the stationary state, which is time-translational invariant. Then,
using the equation of motion, it is immediate to verify that $\dot
C(t) = -A C(t)$, with initial condition given by the covariance matrix
between degrees of freedom at equal time: $C(0) = \sigma$. The
corresponding solution is:
\begin{equation}
C(t) = \exp(-A t) \sigma
\end{equation}
Note that, in general $\sigma$ and $A$ do not commute. It
is also straightforward to recover the response function
$R(t) = \exp(-At)$, since the GFR imposes the following equation
to hold:
\begin{equation} \label{GC}
R(t) = C(t)\sigma^{-1}.
\end{equation}
In general
this function can be written as:
\begin{equation} \label{twoexp}
 \sum_\alpha R_\alpha \exp(-\lambda_\alpha t)
\end{equation}
where $R_\alpha$ are constant matrices, and $\lambda_\alpha$ are the eigenvalues of $A$.
The $i,j$ element of the matrix $R(t)$ is the response function
$\vR_{X_i,X_j}(t)$ for the corresponding degrees of freedom. 
However, at
odds with EFR and KFR, this quantity cannot be expressed
in terms of the correlation $C_{X_i,X_j}(t)$ only, since in general
all the degrees of freedom are coupled:
\begin{equation}
\vR_{X_iX_j}(t)=\sum_k(\sigma^{-1})_{kj}C_{X_iX_k}(t),
\end{equation}
which appears as a violation of EFR or KFR, even if GFR is still valid. 

In the following, we will explicitly walk through this analysis in
two different limit conditions: 
\begin{enumerate}
	\item the free case, when the harmonic force $kx$ can be neglected;
	\item the overdamped case, when inertia $m \ddot{x}$ can be neglected.
\end{enumerate}

\subsection{Dynamics of the free particle \label{section3.1}}

In the limit $k=0$, and setting $m=1$ without loss of generality,
equation (\ref{3mainequation}) becomes
\begin{equation}
\dot{v}=-\gamma_{f}v-\frac{\gamma_{s}}{\tau_{s}}\int^{t}_{-\infty}e^{-\frac{t-t'}{\tau_{s}}}v(t')dt'+\rho_{f}(t)+\rho_{s}(t),\label{3kuboequation}
\end{equation}
where we have introduced velocity $v\equiv dx/dt$.

Eq.~\eqref{3kuboequation} can be mapped to (\ref{2main}) by:
\begin{equation}
\left(\begin{array}{c}
\dot{v} \\ 
\dot{u}
\end{array}\right)
=-\left(\begin{array}{cc} \gamma_{f} & \gamma_{s} \\ 
-\frac{1}{\tau_s} & \frac{1}{\tau_{s}}
\end{array} \right)\left(\begin{array}{c}
v \\ 
u
\end{array}\right)+\left(\begin{array}{c}\sqrt{2 \gamma_{f} T_{f}}\phi_{1}(t)\\ \frac 1{\tau_s\gamma_s} \sqrt{2 \gamma_s T_s}\phi_{2}(t)\end{array} \right) .
\end{equation}
where $\phi_{1}(t)$ and $\phi_{2}(t)$ are independent normalized
white noises, and $u(t)$ is an auxiliary variable:
\begin{equation}
u(t) = \frac 1{\tau_s}\int_{-\infty}^t e^{-\frac{t-t'}{\tau_s}}\left(v(t')+\sqrt{\frac{2 T_s}{\gamma_s}}\phi_2(t'))\right)dt'.
\end{equation}

Denoting $\alpha=\gamma_f/\gamma_s$, $\nu=\tau_s \gamma_s$,
$\zeta^{-1} = (1+\alpha)(1+\nu \alpha)$ and $\Delta T = T_s - T_f$, a
straightforward calculation of covariance matrix gives:
\begin{equation}
\sigma=\left(\begin{array}{cc} T_f + \zeta \Delta T&-\alpha \zeta \Delta T \\- \alpha \zeta \Delta T& T_s/\nu - \alpha \zeta \Delta T\end{array} \right).  \label{3sigmakubo}
\end{equation}
The more general formula for the response as a function of
correlations is given by the GFR, Eq.~\eqref{ger}:
\begin{equation}\label{kubo_correct}
\frac{\overline{\delta v(t)}}{\delta v(0)}=\sigma^{-1}_{11}\left<v(t)v(0)\right>+\sigma^{-1}_{12}\left<v(t)u(0)\right>.
\end{equation}
We can observe two different scenarios: in the case $T_{f}=T_{s}\equiv T$, i.e. $\Delta T= 0$, $\sigma$ is diagonal with $\sigma_{11}=T$ and $\sigma_{22}=T/\nu$,
condition (\ref{3equilibriumfdt}) is restored and, independently by
the values of other parameters, a direct proportionality between
$C_{vv}$ and $R_{vv}$ is obtained (EFR). This is not the only case for
this to happen: e.g. for fixed $\nu = \tau_s\gamma_s$, one has two
possible limits:
\begin{equation}
\begin{array}{ccc}\sigma \approx \left(\begin{array}{cc}T_s &0 \\0& T_s/\nu\end{array} \right) & \mbox{or} & \sigma\approx \left(\begin{array}{cc}T_f &0 \\0& T_s/\nu\end{array} \right),
\end{array}
\end{equation}
respectively for $\gamma_f \ll \gamma_s$, i.e. $\alpha\rightarrow 0$, or
$\gamma_f \gg\gamma_s$, i.e. $\alpha\rightarrow\infty$.  More in
general, when $T_{f}\neq T_{s}$, the coupling term $\sigma_{12}$
differs from zero and a ``violation'' of EFR emerges between the
coupling of different degrees of freedom. 

\begin{figure}
\includegraphics[width=10cm,clip=true]{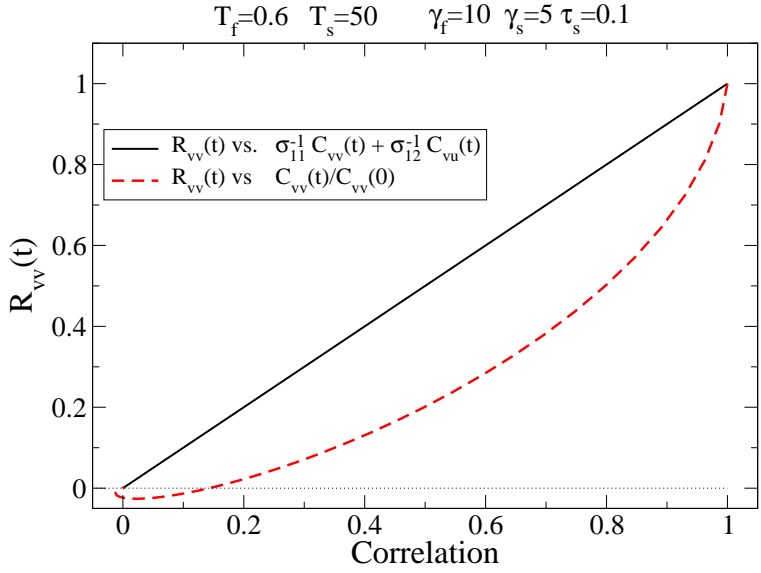}
\caption{Free particle with viscosity and memory, whose dynamics is
given by Eq.~\eqref{3kuboequation}: here we show the parametric plot
of the velocity response to an impulsive perturbation at time $0$,
versus two different correlations. The Einstein relation, which is not
satisfied, would correspond to a linear shape with slope $1$ for the
red dashed line. The black line shows that the GFR holds.
\label{fig:kubo}}
\end{figure}

The situation is clarified by Fig.~\ref{fig:kubo}, where EFR is
violated and the GFR holds: response $R_{vv}(t)$, when plotted against
$C_{vv}(t)$, shows a non-linear and non-monotonic graph. Anyway a
simple linear plot is restored when the response is plotted against
the linear combination of correlations indicated by
formula~\eqref{kubo_correct}. In this case it is evident that the
``violation'' cannot be interpreted by means of any effective
temperature: on the contrary it is a consequence of having ``missed''
the coupling between variables $v$ and $u$, which gives an additive
contribution to the response of $v$.

It is interesting to note that, even when $T_f=T_s$
(i.e. $\sigma_{12}=\sigma_{21}=0$), $C_{vv}(t) \propto R_{vv}(t)$ is a
linear combination of two different exponentials, see
Eq.~\eqref{twoexp}, then its derivative is in general not proportional
to $R_{vv}$, i.e. the KFR does not hold. An example of this situation
will be discussed in Section~\ref{sec:granu} and
Figure~\ref{fig:granususc}.

The above consideration can be easily generalized to the case where
several slow thermostats are present:
let us suppose that there are $N-1$ thermostats at temperature $T$
(the fast one and $N-2$ slow ones) and one at temperature $T_{1}$. In
this case it is possible to show that the off-diagonal terms in $\sigma$ are
proportional to $(T-T_{1})$.

It is useful to stress the role of Markovianity, which is relevant for
a correct prediction of the response. In fact the marginal probability
distribution of velocity $P_{m}(v)$ can be computed straightforward
from (\ref{3kuboequation}) and has always a Gaussian shape. By that,
one could be tempted to conclude, inserting $P_{m}(v)$ inside
GFR, that proportionality between response and
correlation holds also if $T_{f}\neq T_{s}$, in contradiction with
(\ref{3sigmakubo}). This conclusion is wrong, as stated at the
beginning of this section, because the process is Markovian only if
both the variable $v$ and the "hidden" variable $u$ are considered.

\subsection{Overdamped limit with harmonic potential\label{section3.2}}

When $k\neq 0$, in the overdamped limit we can neglect the left-hand
side term in (\ref{3mainequation}), and the equation, after an
integration by parts, reads:
\begin{equation}
\gamma_f \dot{x}=-\left(k + \frac{\gamma_s}{\tau_s}\right) x + \frac{\gamma_s}{\tau_s^2}\int^{t}_{-\infty}e^{-\frac{t-t'}{\tau_{s}}}x(t')dt'+\rho_{f}(t)+\rho_{s}(t).\label{overdamped}
\end{equation}
This model has been discussed in the context of driven glassy
systems~\cite{CK00,ZBCK05}.

In order to restore Markovianity, we can map this equation in:
\begin{equation} \label{3overdampedequations}
\left(\begin{array}{c}
\dot{x} \\ 
\dot{u}
\end{array}\right)= -
\left(
\begin{array}{cc}
\frac k{\gamma_f}+\frac{\gamma_s}{\gamma_f \tau_s} & -\frac{\gamma_s}{\gamma_f \tau_s}\\
-\frac 1{\tau_s} & \frac 1{\tau_s}
\end{array}
\right)
\left(\begin{array}{c}
x \\ 
u
\end{array}\right)
+\left(\begin{array}{c}\sqrt{\frac{2 T_f}{\gamma_f}}\phi_1 \\
\sqrt{\frac{2 T_s}{\gamma_s}}\phi_2
\end{array}\right)
\end{equation}
where, as before, $\phi_1$ and $\phi_2$ are independent white noises, while the auxiliary variable $u$ now reads:
\begin{equation}
u(t) = \frac 1{\tau_s}\int_{-\infty}^t e^{-\frac{t-t'}{\tau_s}}\left(x(t')+\tau_s\sqrt{\frac{2 T_s}{\gamma_s}}\phi_2(t'))\right)dt'.
\end{equation}

\begin{figure}
\includegraphics[width=10cm,clip=true]{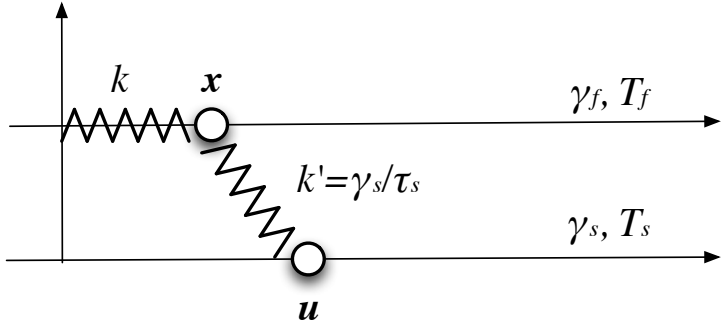}
\caption{Sketch of a simple model which is described by Equation~\eqref{3overdampedequations}. 
\label{fig:molle}}
\end{figure}

Equation~\eqref{3overdampedequations} describes the overdamped
dynamics of the model depicted in Fig.~\ref{fig:molle}: a first
particle at position $x$ is coupled to a thermostat with viscosity
$\gamma_f$ and temperature $T_f$, and to the origin by a spring of
elastic constant $k$; a second particle at position $u$ is coupled to
a thermostat with viscosity $\gamma_s$ and temperature $T_s$; the
coupling between the two particles is a spring of elastic constant
$k'=\gamma_s/\tau_s$. The first particle, when uncoupled from the
second particle, has a characteristic time $\tau_f=\gamma_f/k$.

Using non-dimensional parameters $\alpha=\gamma_f/\gamma_s$ and
$\eta=\tau_s/\tau_f$, the covariance matrix $\sigma$ reads
\begin{equation}
\sigma=\left(\begin{array}{cc} 
\frac{T_f}{k} + \frac{1}{\alpha\Gamma}\frac{\Delta T}k
& 
\frac{T_s}{k} - \frac{1}{\Gamma}\frac{\Delta T}k
\\
\frac{T_s}{k} - \frac{1}{\Gamma}\frac{\Delta T}k
& 
\left(1+\frac{k}{k'}\right)\frac{T_s}k  - \frac{1}{\Gamma}\frac{\Delta T}{k}
\end{array} \right). 
\end{equation}
where $\Gamma =  1+ \frac{1}{\alpha} + \eta$ and $\Delta T = T_s-T_f$.
As we can see, a diagonal form for the matrix $\sigma$ is not
recovered, even for $\Delta T=0$, i.e. for this model the EFR 
never holds.

Let us consider now the KFR, with a force field $h$ coupled to the
variable $x$. First, we note that, from~\eqref{GC},
\begin{equation}
\dot{C}=-RA\sigma
\end{equation}
(where commutativity between $R$ and $A$ has been used). Therefore,
in general, since $\frac{\overline{\delta x}(t)}{\delta h(0)}=R_{xx}/\gamma_f$, one has
\begin{equation} \label{2contr}
\frac{\overline{\delta x}(t)}{\delta h(0)}=-\frac{(A\sigma)^{-1}_{xx}}{\gamma_f}\dot C_{xx}-\frac{(A\sigma)^{-1}_{ux}}{\gamma_f}\dot C_{xu}.
\end{equation}
Then, it is easy to see that the condition to have KFR is
$(A\sigma)_{ux}=0$. Here the matrix $A\sigma$ reads:
\begin{equation}
A\sigma=\left(\begin{array}{cc} 
\frac{1}{\tau_f}\frac{T_f}{k}
& 
-\frac{1}{\tau_f \Gamma}\frac{\Delta T}{k}
\\
\frac{1}{\tau_f \Gamma}\frac{\Delta T}{k}
&
\frac{1}{\tau_s}\frac{T_s}{k'}
\end{array} \right),
\end{equation}
and the condition $(A\sigma)_{ux}=0$ is equivalent to $\Delta T = 0$.


\begin{figure}
\includegraphics[width=10cm,clip=true]{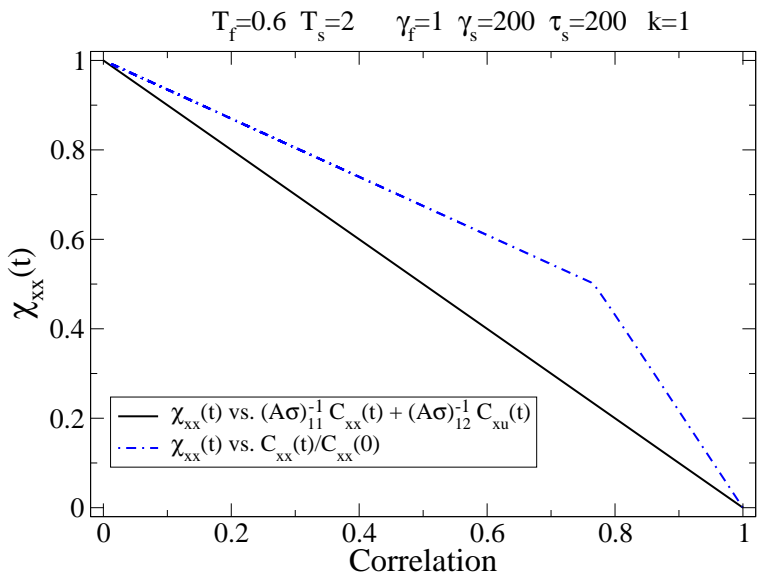}
\caption{Overdamped motion of a particle with harmonic potential, viscosity and memory, whose dynamics is described by Eq.~\eqref{3overdampedequations}: we show here the parametric plot of integrated response versus two different self-correlations. The Kubo formula KFR, which is not satisfied, would correspond to a straigth line of slope $-1$ for the blue dashed curve. The black line shows that the GFR holds. \label{fig:ovmem_restore}}
\end{figure}
As observed in Fig.~\ref{fig:ovmem_restore}, in analogy with the
previous Fig.~\ref{fig:kubo}, the GFR always holds (solid lines),
while the KFR is not verified because it ignores the coupling between
relevant degrees of freedom. In general the local
slope $-s(t)$ of the parametric curve $\chi_{xx}(t)$ vs. $C_{xx}(t)$
is given by
\begin{equation}
s(t)=-\frac{d \chi_{xx}}{d C_{xx}}=-\frac{d\chi(t)}{dt}\left[\frac{dC_{xx}(t)}{dt}\right]^{-1}
\end{equation}
which, for the model in Eq.~\eqref{3overdampedequations}, has two
different limits:
\begin{eqnarray}
1/s(t) &\to T_f \;\;\;\;\;\;\; &\textrm{for} \;\; t \to 0,\\
1/s(t) &\to 1/s_{\infty} =T_s-K \Delta T \;\;\;\;\;\;\; &\textrm{for} \;\;  t \to \infty,
\end{eqnarray}
where
\begin{equation} \label{monster}
K=\frac{2+\Gamma-\sqrt{\Gamma^2-4\eta}}{2\Gamma}.
\end{equation}

Actually, as observed in~\cite{CK00,ZBCK05}, when $\tau_s \gg \tau_f$
the parametric plot $\chi_{xx}(t)$ vs. $C_{xx}(t)$ takes the form of a broken line with two slopes:
the point where the slope abruptly changes, corresponds to the
intermediate plateau of $C_{xx}(t)$ (i.e. when $\tau_s \gg t \gg
\tau_f$) and is located at a position on the $\chi$-axis $\sim y_0
\chi(\infty)$ with $y_0= \frac{T_f}{T_s}\frac{k'}{k}$ : it becomes
visible in the plot if $y_0 \sim\frac{1}{2}$ (i.e. it is not close to
$1$ or $0$). In this case, since $K \to 0$, the two slopes are $1/T_f$
and $1/T_s$. This observation has driven a series of real and
numerical experiments where the parametric plot $\chi_{xx}(t)$
vs. $C_{xx}(t)$ (or their Fourier transforms for the
frequency-dependent susceptibilities) were measured for some degree of
freedom in slowly driven~\cite{BB02} or aging~\cite{HO02,MLDR08}
glassy systems, including models for densely packed granular materials~\cite{S02}. In the next section we discuss this limit and other
interesting cases from the point of view of the GFR: this can be
useful to understand when the KFR-inspired parametric plot is
meaningful and why.

\subsection{Phenomenology of the $\chi$ vs. $C$ parametric plot\label{section3.3}}

In this section we probe the hypothesis that the relative relevance of
the contributions $\frac{(A\sigma)^{-1}_{ux}}{\gamma_f}\dot C_{xu}(t)$
and $\frac{(A\sigma)^{-1}_{xx}}{\gamma_f}\dot C_{xx}(t)$ to the
response $\frac{\overline{\delta x}(t)}{\delta h(0)}$ depends on the time-scale of observation. In
particular we consider the time-integrals of these two contributions,
such that $\chi_{xx}(t)=Q_{xx}(t)+Q_{xu}(t)$, see Eq.~\eqref{2contr}:
\begin{eqnarray}
Q_{xx}(t)&=\frac{(A\sigma)^{-1}_{xx}}{\gamma_f}[C_{xx}(0)-C_{xx}(t)]\\
Q_{xu}(t)&=\frac{(A\sigma)^{-1}_{ux}}{\gamma_f}[C_{xu}(0)-C_{xu}(t)].
\end{eqnarray}

The two eigenvalues of the matrix $A$, determining the time-scales of
the system $1/\lambda_+$ and $1/\lambda_-$, read:
\begin{equation} \label{eigen}
\lambda_{\pm}=\frac{1}{2\tau_s}\left[ \Gamma \pm \sqrt{\Gamma^2-4\eta} \right].
\end{equation}
As suggested by the interpretation given in Fig.~\ref{fig:molle}, the
parameters $\tau_f=\gamma_f/k$ and $\tau_s$ should act as time-scales
when they are separated enough. Indeed, an inspection of
formula~\eqref{eigen} shows that the inverse of $\lambda_+$ and
$\lambda_-$ are proportional to $\tau_f$ and $\tau_s$ when they are well separated, i.e. $\tau_s
\gg \tau_f$ or $\tau_f \gg \tau_s$. However this is a limit case, and more general conditions
can be considered.

\begin{table}
\begin{tabular}{|c|||c|c||c|c||c|c||c|c||c|c|}\hline
case &$T_s$ &$T_f$  &$\gamma_s$ &$\gamma_f$ &$\tau_s$ &$\tau_f$           &k   &$k'$                 &$\frac{1}{\lambda_-}$ &$\frac{1}{\lambda_+}$\\\hline
a    &5     &0.2    &20         &40         &30       &20                 &2   &2/3                  &47.3        &12.7\\
b    &2     &0.6    &200        &1          &200      &1                  &1   &1                    &400         &0.5 \\
c    &2     &0.6    &100        &100        &2        &1000               &0.1 &50                   &2000        &1\\
d    &10    &2      &1          &50         &10       &50                 &1   &0.1                  &51.2        &9.76\\\hline 
\end{tabular}
\caption{Table of parameters for the $4$ cases presented in
Figures~\ref{fig:viol} and~\ref{fig:contrib}. The effective time of
the ``fast'' bath is defined as $\tau_f=\gamma_f/k$, while the
effective spring constant coupling $x$ with $u$ is defined as
$k'=\gamma_s/\tau_s$.\label{paramtab}}
\end{table}

Our choices of parameters, always with $T_f \neq T_s$, are resumed in
Table~\ref{paramtab}: a case (a) where the time-scales are mixed, and
three cases (b), (c) and (d) where scales are well separated. In
particular, in cases (c) and (d), the position of the intermediate
plateau is shifted at one of the extremes of the parametric plot,
i.e. only one range of time-scales is visible.  Of course we do not
intend to exhaust all the possibilities of this rich model, but to
offer a few examples which are interesting for
the following question: what is the meaning of the usual
``incomplete'' parametric plot $\chi_{xx}$ versus $C_{xx}$, which
neglects the contribution of $Q_{xu}$?

\begin{figure}
\includegraphics[width=12cm,clip=true]{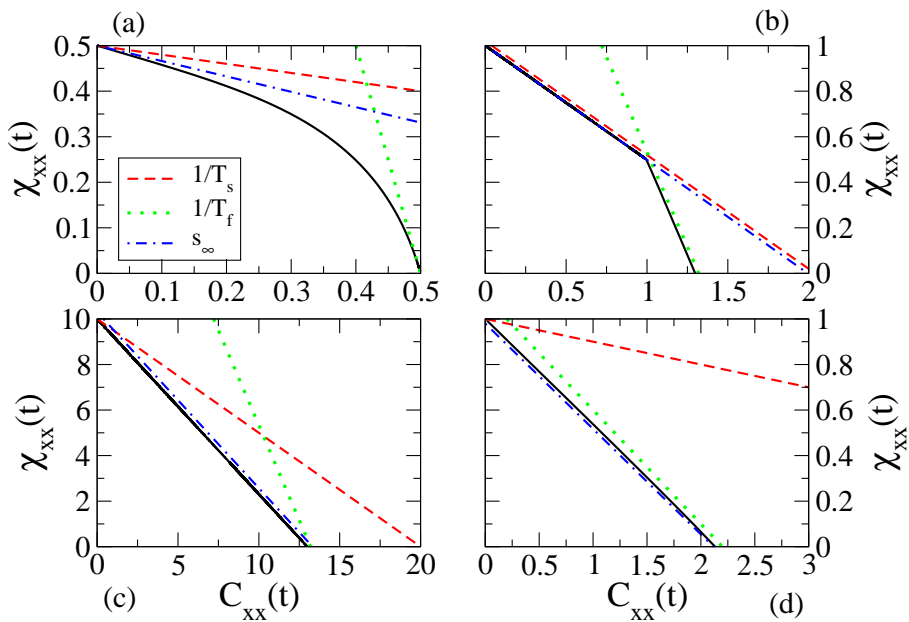}
\caption{Parametric plots of integrated response $\chi_{xx}(t)$ versus
self-correlation $C_{xx}(t)$ for the model in
Eq.~\eqref{3overdampedequations} with parameters given in
Table~\ref{paramtab}. Lines with slopes equal to $1/T_s$, $1/T_f$ and
$s_\infty$ are also shown for reference. \label{fig:viol}}
\end{figure}

\begin{figure}
\includegraphics[width=12cm,clip=true]{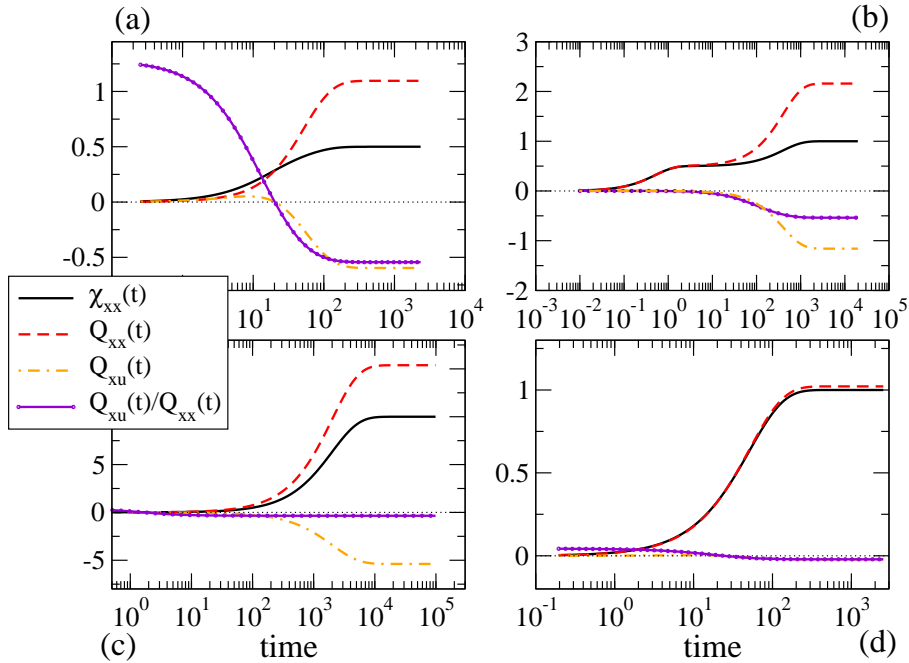}
\caption{Integrated response $\chi_{xx}(t)$ as a function of time, for the model in
  Eq.~\eqref{3overdampedequations} with parameters given in
  Table~\ref{paramtab}. The curves $Q_{xx}(t)$ and $Q_{xu}(t)$,
  representing the two contributions to the response, i.e.
  $\chi_{xx}(t)=Q_{xx}(t)+Q_{xu}(t)$, are also shown. The violet curve
with small circles represents the ratio $Q_{xu}(t)/Q_{xx}(t)$. \label{fig:contrib}}
\end{figure}

The parametric plots, for the cases of Table~\ref{paramtab}, are shown
in Figure~\ref{fig:viol}. In Figure~\ref{fig:contrib}, we present the
corresponding contributions $Q_{xx}(t)$ and $Q_{xu}(t)$ as
functions of time. We briefly discuss the four cases:
\begin{enumerate}

\item[(a)] If the timescales are not separated, the general form of
the parametric plot, see Fig.~\ref{fig:viol}a, is a curve. In fact, as
shown in Fig.~\ref{fig:contrib}a, the cross term $Q_{xu}(t)$ is
relevant at all the time-scales. The slopes at
the extremes of the parametric plot, which can be hard to measure in
an experiment, are $1/T_f$ and $s_\infty \neq 1/T_s$. Apart from that,
the main information of the parametric plot is to point out the
relevance of the coupling of $x$ with the ``hidden'' variable $u$.

\item[(b)] In the ``glassy'' limit $\tau_s \gg \tau_f$, with the
constraint $y_0=\frac{T_f}{T_s}\frac{k'}{k}\sim 1/2$, the well known broken line is found, see
Fig.~\ref{fig:viol}b, as discussed at the end of the previous
section. Figure~\ref{fig:contrib}b shows that $Q_{xu}(t)$ is
negligible during the first transient, up to the first plateau of
$\chi(t)$, while it becomes relevant during the second rise of
$\chi(t)$ toward the final plateau. 

\item[(c)] If $\tau_f \gg \tau_s$, the parametric plot,
Fig.~\ref{fig:viol}c, suggests an equilibrium-like behavior (similar
to what one expects for $T_f=T_s$) with an effective temperature
$1/s_{\infty}$ which is different from both $T_f$ and $T_s$. Indeed,
this case is quite interesting: the term $Q_{xu}$ is of the same order
of $Q_{xx}$ during all relevant time-scales, but $Q_{xu}/Q_{xx}$ appears to be
almost constant. This leads to observe a KFR-like plot with a
non-trivial slope. The close similarity between $Q_{xx}$ and $Q_{xu}$
is due to the high value of the coupling constant
$k'=\gamma_s/\tau_s$.

\item[(d)] In the last case, always with $\tau_f \gg \tau_s$, the
contribution of $Q_{xu}(t)$ is negligible at all relevant time-scales
(see Fig.~\ref{fig:contrib}), giving place to a straigth parametric
plot, shown in Fig.~\ref{fig:viol}, with slope $1/T_f$. The low value of the
coupling constant $k'$ is in agreement with this observation.

\end{enumerate}


The lesson learnt from this brief study is that the shape of the
parametric plot depends upon the timescales and the relative coupling
$k'/k$. This is consistent with the fact that the correct formula for
the response is always the GFR:
$\frac{\overline{\delta x}(t)}{\delta h(0)}=\dot{Q}_{xx}+\dot{Q}_{xu}$. However, the definition of an
effective temperature through the relation $T_{eff}(t)\frac{\overline{\delta x}(t)}{\delta h(0)}=\dot{Q}_{xx}(t)$ in general (see
case a), does not seem really useful. In particular limits, the
behavior of the additional term $Q_{xu}$ is such that $R \propto
\dot{Q}_{xx}$ in a range of time-scales, and therefore the measure of
$T_{eff}$ becomes meaningful.

\section{Granular gases} \label{sec:granu}
In the previous Section we have shown in the analytically tractable
  case of linear Langevin equations how the presence of coupling among
  different degrees of freedom plays a role in the specific form of
  the GFR, which is in general different from the EFR and KFR. Such a
  feature is certainly not specific of the considered model: the
  non-equilibrium dynamics of a many particles system may present the
  same kind of nontrivial correlations among degrees of freedom,
  usually due to strong inhomogeneities generated by the lack of
  conservation laws valid at equilibrium. In these models the
  stationary distribution is not known and the use of the GFR for
  response analysis is rather subtle.\\ In the following we shall
  analyze granular gases in the steady state, which offer an
  interesting benchmark of this ideas.

\subsection{The model}
Let us consider a $d$-dimensional model for driven granular gases~\cite{PLMPV98,PLMV99,WM96,NETP99}: $N$
identical disks (in $d=2$) or rods (in $d=1$) of diameter $1$ in a
volume $V=L\times L$ (in $d=2$) or total length $L$ (in $d=1$) with
inelastic hard core interactions characterized by an instantaneous
velocity change
\begin{equation}
\bv_i'=\bv_i-\frac{1+r}{2}[(\bv_i-\bv_j)\cdot \hat{n}]\hat{n},
\end{equation}
where $i$ and $j$ are the label of the colliding particles, $\bv$ and
$\bv'$ are the velocity before and after the collision respectively,
$\hat{n}$ is the unit vector joining the centers of particles and
$r \in [0,1]$ is the restitution coefficient which is equal to
$1$ in the elastic case. Each particle $i$ is coupled to a ``thermal
bath'', such that its dynamics (between two successive collisions)
obeys
\begin{equation}
\frac{d\bv_i}{dt}=-\frac{1}{\tau_b} \bv_i + \sqrt{\frac{2T_b}{\tau_b}} \eta_i(t)
\end{equation}
where $\tau_b$ and $T_b$ are parameters of the ``bath'' and
$\eta_i(t)$ are independent normalized white noises. We restrict
ourselves to the dilute or liquid-like regime, excluding more dense
systems where the slowness of relaxation prevents clear measures and
poses doubts about the stationarity of the regime and its ergodicity: in
practice we consider packing fractions (fraction of occupied volume)
$\psi=N/(4V)$ in the range $0.01 \div 0.5$. Two important observables of the
system are the mean free time between collisions, $\tau_c$, and the
so-called granular temperature $T_g=\langle |\bv|^2 \rangle/d$. 
\\In this model is possible to recover 
two different regimes. When the thermostat is dominant,
i.e. when $\alpha=\tau_c/\tau_b \gg 1$, grains thermalize, on average, with 
the bath before experiencing a collision and the inelastic effects are negligible. 
This is an ``equilibrium-like'' regime, similar to the elastic case $r=1$, where 
the granular gas is spatially homogeneous, the distribution of velocity is Maxwellian and $T_g=T_b$. On the
contrary, when $\tau_{c}<\tau_{b}$, non-equilibrium effects can emerge such 
as deviation from Maxwell-Boltzmann statistics, spatial inhomogeneities and $T_g < T_b$~\cite{PLMPV98,PLMV99,WM96,NETP99}.
This ``granular regime'', easily reached when packing fraction is increased or inelasticity is reduced,
is characterized by correlations among different particles. This peculiarity is the key 
ingredient for a correct response analysis of these systems, as we shall see in the following.

\subsection{Failure of the EFR for strong dissipation}
An analysis of the FDR for the previous model has been performed
in~\cite{PBL02,PBV07} (in $d=2$) and~\cite{VPV08} (in $d=1$), and
discussed also in~\cite{BPRV08}. Similar results are also obtained for
other models, such as the inelastic Maxwell model on a $d=2$
lattice driven by a Gaussian thermostat~ and mean field granular gases~\cite{BPV08,BSL07}. 
The protocol used in numerical experiments cited above is the following:

\begin{enumerate}

\item the gas is prepared in a ``thermal'' state, with random
  velocity components extracted from a Gaussian with zero average and
  given variance, and positions of the particles  chosen uniformly
  random in the box, avoiding overlapping configurations.

\item The system is let evolve until a
statistically stationary state is reached, which is set as time $0$:
we verify that the total kinetic energy fluctuates around an average 
value which does not depend on initial
conditions. 

\item A copy of the system is obtained, identical to the
original but for one particle, whose $x$ (for instance) velocity component
is incremented of a fixed amount $\delta v(0)$. 

\item Both systems are let evolve with the unperturbed
  dynamics. For the random thermostats, the same noise realization is used. The
  perturbed tracer has velocity $v'(t)$, while the unperturbed one has
  velocity $v(t)$, so that $\delta v(t)=v'(t)-v(t)$.

\item After a time
$t_{max}$ large enough to have lost memory of the configuration at
time $0$, a new copy is done with perturbing a new random particle and
the new response is measured. This procedure is repeated until a sufficient collection of data is obtained

\item Finally the autocorrelation function
  $C_{vv}(t)=\langle v(t)v(0) \rangle$ in the original system and the response $R_{vv}(t)\equiv\frac{\overline{\delta
      v}(t)}{\delta v(0)}$ are measured. 

\end{enumerate}

\begin{figure}
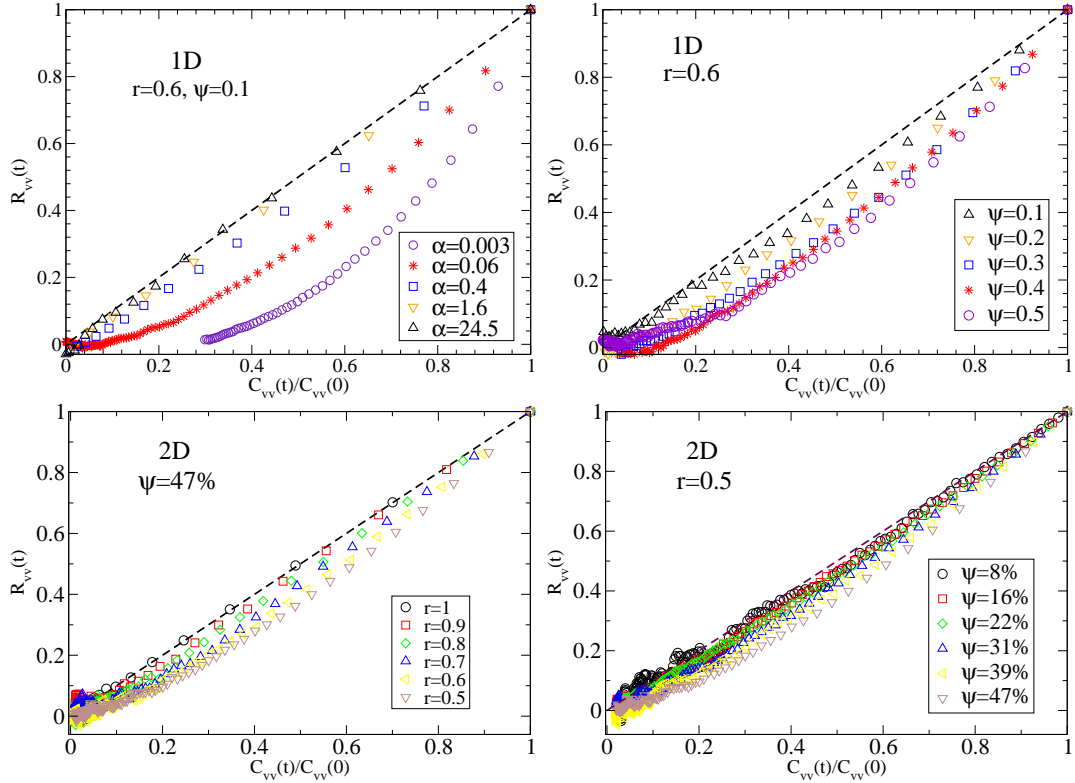

\includegraphics[width=7cm,clip=true]{1d_parametric_alpha.eps}
\includegraphics[width=7cm,clip=true]{1d_parametric_phi.eps}\\
\includegraphics[width=7cm,clip=true]{2d_parametric_r.eps}
\includegraphics[width=7cm,clip=true]{2d_parametric_phi.eps}
\caption{Parametric plots to check the EFR, for $d=1$ and $d=2$ models of inelastic
hard-core gases with thermal bath. Different choices of parameters $r$
(restitution coefficient), $\alpha=\tau_c/\tau_b$ and $\psi$ (packing
fraction) are shown: note that one can change $\alpha$ at $\psi$ or
$r$ fixed (changing $\tau_b$), but - in general - changes in $\psi$ or
$r$ determine also changes in $\alpha$ (because of changes in
$\tau_c$). In all plots, the dashed line marks the Einstein relation
$R_{vv}=C_{vv}(t)/C_{vv}(0)$. \label{fig:granuviol}}
\end{figure}

\begin{figure}
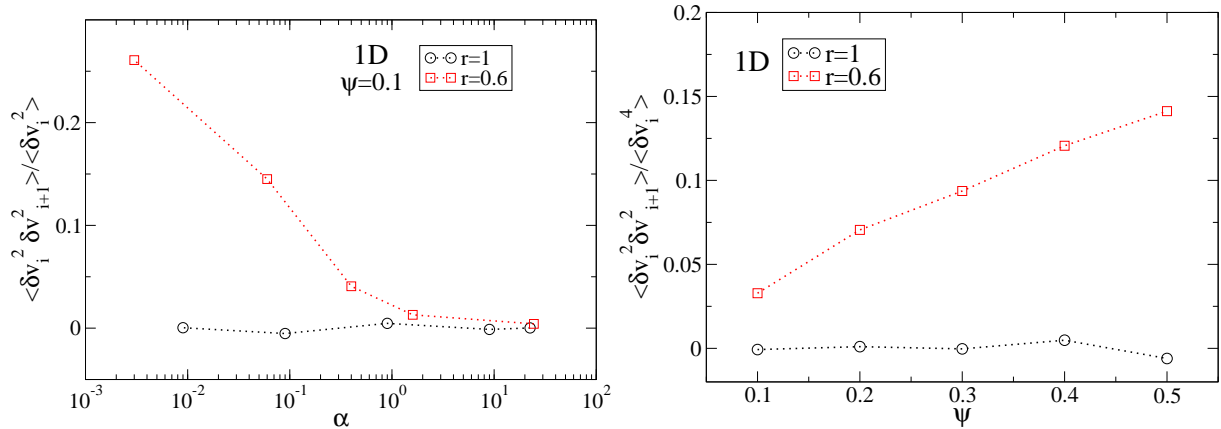

\includegraphics[width=8cm,clip=true]{1d_correlations_alpha.eps}
\includegraphics[width=8cm,clip=true]{1d_correlations_phi.eps}
\caption{Static (same time) particle-particle correlations for energy
fluctuations, for the $d=1$ inelastic hard rods gas. In the inelastic case (squares) the coefficient
is higher when the dissipation gets stronger, i.e. for small $\alpha$ (left) or high packing fraction (right). The coefficient 
in the elastic case is also shown, which is negligible for all the values of the parameters. \label{fig:granucorr}}
\end{figure}

The main result of those studies
can be so summarized: in the dilute limit (where the packing fraction $\psi \to 0$) or in the elastic limit ($r \to
1$), or in the limit of efficient thermostat $\alpha \to \infty$
(which is usually implied by the dilute limit), the Einstein relation,
Eq.~\eqref{er}, is recovered for the velocity of a tracer particle, i.e. $R_{vv}(t)=C_{vv}(t)/T_g$. On the contrary,
when these conditions fails one can observe strong deviations from EFR. \\
In addiction, there are same remarkable points:
\begin{itemize}
 \item Non-Gaussian velocity distributions can appear also in the dilute regime, but they seems to have a minor role
in the violations of EFR~\cite{PBV07}.
\item The same scenario holds in dimension $d=1$, where the tracer is sub-diffusive,
i.e. where $\langle |x(t)-x(0)|^2 \rangle \sim t^{1/2}$ which also
implies a non-monotonic $C_{vv}(t)$ with a power-law tail $C_{vv}(t)
\sim -t^{-3/2}$ for large $t$~\cite{VPV08}.
\item When a mixture of two different kinds
of grains (e.g. with different masses or different restitution
coefficients) is considered~\cite{BLP04}, the two components bear
different granular temperatures and this lead to separated FDR in the
dilute limit, i.e. a tracer satisfies the EFR with its own
temperature, making very difficult to obtain a neutral thermometer
based on FDR.
\end{itemize}
The violation of the EFR is more and more pronounced as the inelasticity increases (lower values of
$r$), the importance of the bath is reduced (lower values of $\alpha$) 
or the packing fraction $\psi$ is increased, as shown in Figure \ref{fig:granuviol}. In correspondence of such
variations of parameters, the correlation between velocities of
adjacent particles is also enhanced, a  phenomena which is ruled out in
equilibrium fluids. This effect can be directly measured in many ways,
a possibility is shown in Figure~\ref{fig:granucorr}. In conclusion, the general lesson is that there is a quite clear
correspondence between violations of the EFR and the appearance of
correlations among different degrees of freedom.


\begin{figure}
\includegraphics[width=14cm,clip=true]{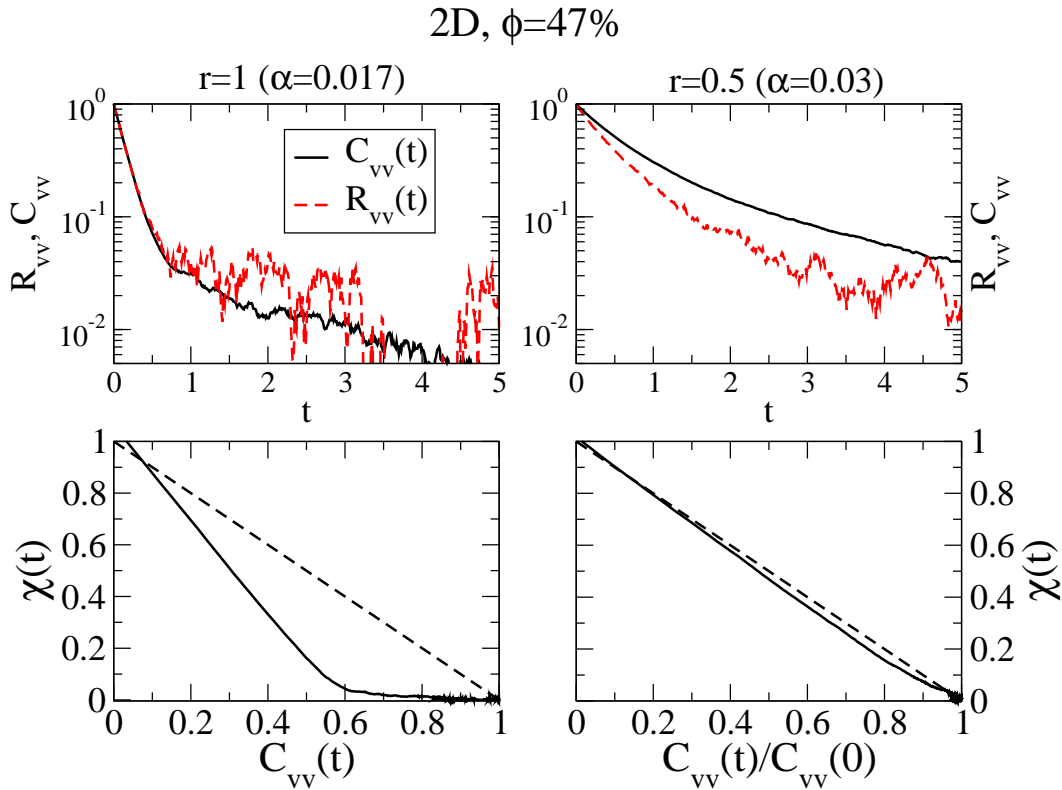}
\caption{Comparison of an elastic (left) and inelastic (right) gas of
hard disks coupled to the thermal bath in $d=2$: the response $R_{vv}$
(denoted as $R$ for simplicity), the susceptibility
$\chi_{vv}(t)=\int_0^t R_{vv}(t') dt'$ and the normalized velocity
auto-correlation $C(t)=C_{vv}(t)/C_{vv}(0)$ are plotted in several
different ways. The top graphs show the validity or breakdown of the
Einstein relation $R=C$ for the elastic or inelastic case
respectively; the bottom graphs display the parametric plot
$\chi_{vv}$ vs. $C_{vv}$ which would follow the form $\chi_{vv}=1-C$
if the second form of FDR, Eq.~\eqref{kubor}, i.e. $R_{vv}=-\dot{C}_{vv}$,
held. \label{fig:granususc}}
\end{figure}

\subsection{A correct prediction of the response}
\begin{figure}
\includegraphics[width=10cm,clip=true]{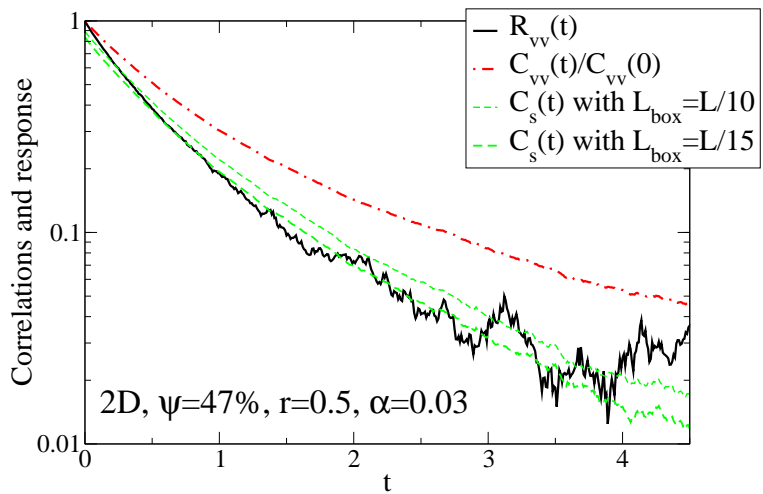}
\caption{Response and correlations of the velocity tracer in the $d=2$
gas of inelastic hard disks for a choice of parameters such that FDR
type (i) (Einstein relation, Eq.~\eqref{er}) is strongly violated. The
dashed green curves show the conjecture $R_{vv}=C_s(t)\equiv\frac{1}{T_g}\langle
v(t)\{v(0)-u[x(0)]\}\rangle$ is probed, where $u(x)$ is the local
velocity field measured by coarse-graining in boxes (centered with the
tracer) of size $L_{box}$. \label{fig:granuseif}}
\end{figure}

Let us begin with an example explaining how a blind comparison between autocorrelation and response
can be misleading. Clearly the EFR is satisfied in the elastic (``equilibrium'') case, 
i.e. $R_{vv}(t) \propto C_{vv}(t)$, thanks to the fact that, 
in the invariant measure, there is a very weak coupling between the tracer velocity and other degrees of freedom. However, it is also interesting to note that the shape of $C_{vv}(t)$ is far from being an exponential, also in the elastic (``equilibrium'') case,
because of the presence of two characteristic times $\tau_b$ and
$\tau_c$. This non-exponential shape of $C_{vv}(t)$ leads to a failure of the KFR, Eq.~\eqref{kuborint}, as put in evidence in
the lower left plot of Fig.~\ref{fig:granususc}. 
It is also instructive to plot the same quantities for a strongly
inelastic case, see right plots of Fig.~\ref{fig:granususc}, where an
almost linear relation seems at work. 
A first rough explanation is the lower granular temperature which is responsible 
for a higher mean free time $\tau_c$, implying that $\tau_c$ and $\tau_b$
 are slightly closer to each other with respect to the elastic case, making the $C_{vv}(t)$
similar to a single exponential.\\
This example show how the only way to have a correct prediction of the response can reside
in the use of the GFR, which is always valid. For a quantitative comparison between correlation functions and response, one needs some hypothesis on the
stationary probability distribution, in particular about the kind of
coupling between different phase-space variables. We report the result
of a simple assumption, partly inspired to an idea of Speck and
Seifert~\cite{ss06}, where correlation among variables is mediated by
a fluctuating ``hydrodynamic'' velocity field $\bu(\bx)$, in such a
way that the relevant part of the stationary probability distribution
for the tracer reads, approximately:
\begin{equation} \label{eq:pdflocal}
P_m(\bv,\bx,t) \sim \exp\left\{-\frac{[\bv-\bu(\bx,t)]^2}{2T_g}\right\},
\end{equation}
with $\bu(\bx,t)$ a local velocity average, defined on a small cell of
diameter $L_{box}$ centered in the particle. This is motivated by the
observation that, at high density or inelasticities, spatially
structured velocity fluctuations appear in the system for some time,
even in the presence of external noise~\cite{NETP99,TPNE01}. The
generalized FDR following from assumption~\eqref{eq:pdflocal} reads
\begin{equation} \label{eq:risplocal}
R_{vv}(t)=C_s(t)=\frac{1}{T_g}\langle v(t) \{v(0)-u[x(0)]\}\rangle.
\end{equation}
and is nicely verified in Fig.~\ref{fig:granuseif}.  In other words,
one has a correction to the ``naive'' expectation $R_{vv}(t)=\langle
v(t)v(0) \rangle/T_g$, i.e. the extra term $-\langle v(t)u[x(0)]
\rangle$ originated by the presence of a ``hydrodynamic'' velocity
field. A recent experiment~\cite{GPCCG09} shows the presence, in a
rather clear way, of a similar extra term (with respect to the KFR)
for a colloidal particle in a toroidal optical trap, in a
non-equilibrium steady state.\\ We conclude this section underlining
the connections between the results obtained for granular gases and Langevin equations. In general, the
behaviour of the Langevin model and of the granular model show some
differences, for example the case expressed in Figure \ref{fig:viol}
(case b) has not a counterpart in these models and the ``effective
temperature'' approach is meaningless, even when times scales are well
separated. However the use of GFR shows how, in both examples, the
response is given by a sum of different contributes, and in some
special limits, the cross correlation term can be neglected and a
comparison between the response and the autocorrelation does make
sense. This happens in the ``equilibrium-like'' cases of the Langevin
model (cfr. Figure \ref{fig:viol}, cases c,d) and in the dilute regime
of the granular gas, in which there is no coupling between the velocity
of the tracer and the ``hidden variable'' embodied by the local
velocity average. On the contrary, when this approximation is not
correct, a response-autocorrelation plot shows strong deviations from
linearity, but can be predicted by taking into account all the
contributes in the computation of the response.

\section{Conclusions}

In this paper we have reviewed different forms of the
fluctuation-dissipation relation for steady states. The most general
one is the GFR, Eq.~\eqref{ger}, which requires the knowledge of the
relevant degrees of freedom and their reciprocal couplings in the
system. When this knowledge is not accessible, the study of the
response to a perturbation of a certain variable, compared to the
correlation of that variable in the unperturbed state, has not a
simple meaning, in general.


As an example, we study in detail two limits of a generalized Langevin
equation with memory, for a particle which moves in a harmonic
potential and is in contact with two thermostats at different
temperatures. In the overdamped case, the response-correlation
parametric plot may reveal a broken line shape, where the two slopes
are given by the inverse temperatures of the baths: a necessary, but
not sufficient, condition for this to happen is that time-scales are
well separated. In the general case, however, the plot can be
more difficult to read, showing intermediate "effective temperatures",
as well as a more general non-linear shape.

Nevertheless, the problem can be recast as a Markovian dynamics, by
means of the introduction of additional degrees of freedoms. The
Markovian dynamics has a stationary distribution and satisfies a general relation
(GFR) between the response and a specific correlation function, which
is related to the stationary measure and takes
into account the coupling between degrees of freedom. We show how this
coupling is responsible for the "violation" of the usual FDR.


An interesting case where correlations play a role in the "violation"
of FDR, is the dynamics of a tracer particle in a driven granular gas.
Here the response is not proportional, in general, to the unperturbed
autocorrelation and an effective temperature does not seem to be
informative, even when the time-scales are well
separated. Nevertheless a GFR should always be valid, provided an
appropriate description of the dynamics is given.  Indeed, as already
stressed by Onsager and Machlup in their seminal work on fluctuations
and irreversible processes~\cite{OM53}, a basic ingredient
for a ``good statistical'' description, is Markovianity. We remind
their important caveat: {\em how do you know you have taken enough
variables, for it to be Markovian?}  We have seen that, following this
suggestion, it is possible to have deeper insight on the so-called
``violations of FDR'' and in some cases (e.g. in
Fig.~\ref{fig:granuseif}), one can try to guess the correct
correlation involved in the dynamics.

\vspace{.5cm}
\noindent {\it Acknowledgments.--} We wish to warmly thank L. Leuzzi, U. Marini Bettolo Marconi, F. Ricci-Tersenghi, L. Rondoni and P. Visco for discussions and a critical reading of the
manuscript. The work of AP is supported by the ``Granular-Chaos'' project, funded by the Italian MIUR under the FIRB-IDEAS grant number RBID08Z9JE.
\section{References}

\bibliographystyle{unsrt}
\bibliography{fluct.bib}

\begin{thebibliography}{10}

\bibitem{E56}
A~Einstein.
\newblock {\em Investigation on the Theory of the Brownian Motion}.
\newblock Dover, 1956.

\bibitem{O31}
L~Onsager.
\newblock Reciprocal relations in irreversible processes. {I}.
\newblock {\em Phys. Rev.}, 37:405, 1931.

\bibitem{O31b}
L~Onsager.
\newblock Reciprocal relations in irreversible processes. {II}.
\newblock {\em Phys. Rev.}, 38:2265, 1931.

\bibitem{K57}
R~Kubo.
\newblock Statistical-mechanical theory of irreversible processes. {I. G}eneral
  theory and simple applications to magnetic and conduction problems.
\newblock {\em J. Phys. Soc. Japan}, 12:570, 1957.

\bibitem{K66}
R~Kubo.
\newblock The fluctuation-dissipation theorem.
\newblock {\em Rep. {P}rog. {P}hys.}, 29:255, 1966.

\bibitem{ECM}
D.~J. Evans, E.~G.~D. Cohen, and G.~P. Morriss.
\newblock Probability of second law violations in shearing steady flows.
\newblock {\em Phys. Rev. Lett.}, 71:2401, 1993.

\bibitem{GC}
G~Gallavotti and E~G~D Cohen.
\newblock Dynamical ensembles in stationary states.
\newblock {\em J. Stat. Phys.}, 80:931, 1995.

\bibitem{CJ}
C~Jarzynski.
\newblock Nonequilibrium equality for free energy differences.
\newblock {\em Phys. Rev. Lett.}, 78:2690, 1997.

\bibitem{BPRV08}
U~Marini~Bettolo Marconi, A~Puglisi, L~Rondoni, and A~Vulpiani.
\newblock Fluctuation-dissipation: Response theory in statistical physics.
\newblock {\em Phys. Rep.}, 461:111, 2008.

\bibitem{BCKM98}
J~P Bouchaud, L~F Cugliandolo, J~Kurchan, and M~Mezard.
\newblock {\em Spin Glasses and Random Fields}.
\newblock World Scientific, 1998.

\bibitem{MPRR98}
E~Marinari, G~Parisi, F~Ricci-Tersenghi, and JJ~Ruiz-Lorenzo.
\newblock Violation of the fluctuation-dissipation theorem in
  finite-dimensional spin glasses.
\newblock {\em J. Phys. A.}, 31:2611, 1998.

\bibitem{LN07}
L~Leuzzi and Th.~M Nieuwenhuizen.
\newblock {\em Thermodynamics of the Glassy State}.
\newblock Taylor {$\&$} Francis, 2007.

\bibitem{a72}
G~S Agarwal.
\newblock Fluctuation-disipation theorems for systems in non-thermal
  equilibrium and applications.
\newblock {\em Z. Physik}, 252:25, 1972.

\bibitem{dh75}
U~Deker and F~Haake.
\newblock Fluctuation-dissipation theorems for classical processes.
\newblock {\em Phys. Rev. A}, 11:2043, 1975.

\bibitem{ht77}
P.~H\"anggi and H.~Thomas.
\newblock Time evolution, correlations and linear response of non-{M}arkov
  processes.
\newblock {\em Z. Physik B}, 26:85, 1977.

\bibitem{ht82}
P.~H\"anggi and H.~Thomas.
\newblock Stochastic processes: Time-evolution, symmetries and linear response.
\newblock {\em Phys. Rep.}, 88:207, 1982.

\bibitem{FIV90}
M~Falcioni, S~Isola, and A~Vulpiani.
\newblock Correlation functions and relaxation properties in chaotic dynamics
  and statistical mechanics.
\newblock {\em Physics {L}etters {A}}, 144:341, 1990.

\bibitem{KTH91}
R~Kubo, M~Toda, and N~Hashitsume.
\newblock {\em Statistical physics II: Nonequilibrium stastical mechanics}.
\newblock Springer, 1991.

\bibitem{CK00}
L~F Cugliandolo and J~Kurchan.
\newblock A scenario for the dynamics in the small entropy production limit.
\newblock {\em J Phys Soc Jpn}, 69:247, 2000.

\bibitem{ZBCK05}
F~Zamponi, F~Bonetto, L~F Cugliandolo, and J~Kurchan.
\newblock A fluctuation theorem for non-equilibrium relazational systems driven
  by external forces.
\newblock {\em J. Stat. Mech.}, page P09013, 2005.

\bibitem{CR03}
A~Crisanti and F~Ritort.
\newblock Violation of the fluctuation-dissipation theorem in glassy systems:
  basic notions and the numerical evidence.
\newblock {\em J. Phys. A}, 36:R181, 2003.

\bibitem{R89}
H~Risken.
\newblock {\em The Fokker-Planck equation: Methods of solution and
  applications}.
\newblock Springer- {V}erlag, Berlin, 1989.

\bibitem{BB02}
L~Berthier and J~L Barrat.
\newblock Shearing a glassy material: Numerical tests of nonequilibrium
  mode-coupling approaches and experimental proposals.
\newblock {\em Phys. Rev. Lett.}, 89:095702, 2002.

\bibitem{HO02}
D~H{\'e}risson and M~Ocio.
\newblock Fluctuation-dissipation ratio of a spin glass in the aging regime.
\newblock {\em Phys. Rev. Lett.}, 88:257202, 2002.

\bibitem{MLDR08}
C~Maggi, R~di~Leonardo, J~C Dyre, and G~Ruocco.
\newblock Direct demonstration of fluctuation-dissipation theorem violation in
  an off-equilibrium colloidal solution.
\newblock {\em arXiv:0812.0740}, 2008.

\bibitem{S02}
M~Sellitto.
\newblock Effective temperature and compactivity of a lattice gas under
  gravity.
\newblock {\em Phys. Rev. E}, 66:042101, 2002.

\bibitem{PLMPV98}
A~Puglisi, V~Loreto, U~M~B Marconi, A~Petri, and A~Vulpiani.
\newblock Clustering and non-gaussian behavior in granular matter.
\newblock {\em Phys. Rev. Lett.}, 81:3848, 1998.

\bibitem{PLMV99}
A~Puglisi, V~Loreto, U~M~B Marconi, and A~Vulpiani.
\newblock A kinetic approach to granular gases.
\newblock {\em Phys. Rev. E}, 59:5582, 1999.

\bibitem{WM96}
D~R~M Williams and F~C MacKintosh.
\newblock Driven granular media in one dimension: Correlations and equation of
  state.
\newblock {\em Phys. Rev. E}, 54:R9, 1996.

\bibitem{NETP99}
T~P~C van Noije, M~H Ernst, E~Trizac, and I~Pagonabarraga.
\newblock Randomly driven granular fluids: Large-scale structure.
\newblock {\em Phys. Rev. E}, 59:4326, 1999.

\bibitem{PBL02}
A~Puglisi, A~Baldassarri, and V~Loreto.
\newblock Fluctuation-dissipation relations in driven granular gases.
\newblock {\em Physical Review E}, 66:061305, 2002.

\bibitem{PBV07}
A~Puglisi, A~Baldassarri, and A~Vulpiani.
\newblock Violations of the {E}instein relation in granular fluids: the role of
  correlations.
\newblock {\em J. Stat. Mech.}, page P08016, 2007.

\bibitem{VPV08}
D~Villamaina, A~Puglisi, and A~Vulpiani.
\newblock The fluctuation-dissipation relation in sub-diffusive systems: the
  case of granular single-file diffusion.
\newblock {\em J. Stat. Mech.}, page L10001, 2008.

\bibitem{BPV08}
A~Baldassarri, A~Puglisi, and A~Vulpiani.
\newblock Fluctuations and response in granular gases: validity and failure of
  {E}instein relation.
\newblock In A~Co, G~L Leal, R~H Colby, and A~J Giacomin, editors, {\em The XV
  International Congress on Rheology: The Society of Rheology 80th Annual
  Meeting}, volume 1027 of {\em AIP Conference Proceedings}, page 911. AIP,
  2008.

\bibitem{BSL07}
G.~Bunin, Y.~Shokef, and D.~Levine.
\newblock {Frequency-dependent fluctuation--dissipation relations in granular
  gases}.
\newblock {\em Phys. Rev. E}, 77:051301, 2008.

\bibitem{BLP04}
A~Barrat, V~Loreto, and A~Puglisi.
\newblock Temperature probes in binary granular gases.
\newblock {\em Physica A}, 334:513, 2004.

\bibitem{ss06}
T~Speck and U~Seifert.
\newblock Restoring a fluctuation-dissipation theorem in a nonequilibrium
  steady state.
\newblock {\em Europhys. Lett.}, 74:391, 2006.

\bibitem{TPNE01}
E~Trizac, I~Pagonabarraga, T~P~C van Noije, and M~H Ernst.
\newblock Randomly driven granular fluids: Collisional statistics and short
  scale structure.
\newblock {\em Phys. Rev. E}, 65:011303, 2001.

\bibitem{GPCCG09}
J~R Gomez-Solano, A~Petrosyan, S~Ciliberto, R~Chetrite, and K~Gawedzki.
\newblock Experimental verification of a modified fluctuation-dissipation
  relation for a micron-sized particle in a non-equilibrium steady state.
\newblock {\em arXiv:0903.1075}, 2009.

\bibitem{OM53}
L~Onsager and S~Machlup.
\newblock Fluctuations and irreversible processes.
\newblock {\em Phys. Rev.}, 91:1505, 1953.

\end{thebibliography}

\end{document}